\documentclass[11pt]{article}
\pdfoutput=1 
\usepackage{jheppub}

\usepackage{amsmath,amssymb,amsbsy,amstext,amsthm,booktabs,simplewick,exscale,relsize,graphicx,amsfonts,upgreek}
\usepackage{multirow,dcolumn,bm,enumerate,url}
\usepackage{color, graphics}
\usepackage{amsmath,amssymb,amsfonts}
\usepackage{verbatim}   
\usepackage{subfigure}  
\usepackage{acronym}
\usepackage{hyperref}
\usepackage{mathrsfs}
\usepackage{multirow}
 \usepackage{cancel}
 \usepackage{url,nicefrac}
  \usepackage{slashed}
 \usepackage{hyperref}

\def\beq{\begin{equation}\begin{aligned}}
\def\eeq{\end{aligned}\end{equation}}
\def\OO{\mathcal{O}}

\begin{document}

\date{\today}

\title{Superheavy Thermal Dark Matter\\ and Primordial Asymmetries}

\author[1]{Joseph Bramante}
\affiliation[1]{Perimeter Institute for Theoretical Physics, Waterloo, Ontario, N2L 2Y5, Canada}
\author[2]{James Unwin}
\affiliation[2]{Department of Physics,  University of Illinois at Chicago, Chicago, IL 60607, USA}

\abstract{
The early universe could feature multiple reheating events, leading to jumps in the visible sector entropy density that dilute both particle asymmetries and the number density of frozen-out states. In fact, late time entropy jumps are usually required in models of Affleck-Dine baryogenesis, which typically produces an initial particle-antiparticle asymmetry that is much too large. An important consequence of late time dilution, is that a smaller dark matter annihilation cross section is needed to obtain the observed dark matter relic density. For cosmologies with high scale baryogenesis, followed by radiation-dominated dark matter freeze-out, we show that the perturbative unitarity mass bound on thermal relic dark matter is relaxed to $10^{10}$ GeV. We proceed to study superheavy asymmetric dark matter models, made possible by a sizable entropy injection after dark matter freeze-out, and identify how the Affleck-Dine mechanism would generate the baryon and dark asymmetries. }

\maketitle


\section{Introduction}

There are two well-motivated possibilities for generating the baryon asymmetry at high temperatures: right-handed neutrino leptogenesis and the Affleck-Dine baryogenesis \cite{Affleck:1984fy,Dine:1995kz,Dine:2003ax}. The Affleck-Dine mechanism utilises the fact that scalar potentials in supersymmetric (SUSY) models have nearly ``flat directions". In the early universe, gauge  invariant combinations of scalar fields that carry an approximately conserved global quantum number (such as baryon $B$ or baryon-minus-lepton $B-L$ number) become initially displaced to large field values. Once the Hubble parameter drops below a given mass scale, the associated scalar field will roll towards its minimum. If the initially displaced $B$-charged scalar fields have baryon number and charge-parity ($CP$) violating potentials, the evolution of the field to its minimum leads to the growth of a large baryon asymmetry.

Notably, Affleck-Dine baryogenesis often leads to a baryon asymmetry much larger than presently observed. Such a large baryon asymmetry must be subsequently diluted, usually through an injection of entropy into the thermal bath and/or strong washout from sphaleron processes during a phase transition. Fortunately such entropy injection events are ubiquitous in UV completions of the Standard Model. In particular string theory typically introduces a large number of gravitationally coupled scalars which decay at late cosmological times, diluting previous particle asymmetries and relic abundances \cite{Banks:1993en,de Carlos:1993jw,Moroi:1999zb}. This motivates serious consideration of the possibility that at some point in cosmological history, there were large dilutions in asymmetries and particle number due to entropy injection. The occurrence of these large entropy dumps can significantly impact what is regarded as a target range for model building when considering the appropriate freeze-out abundance of dark matter or the magnitude of particle-antiparticle asymmetries. 

Griest and Kamionkowski \cite{Griest:1989wd} argued that if the dark matter is ever in thermal equilibrium with the Standard Model bath, and its freeze-out annihilation cross section is required to be perturbative, then this restricts the dark matter mass to be $m_{\rm DM}\lesssim100$ TeV. An important caveat to this conclusion is that subsequent entropy production can dilute the abundance of frozen out states. Here we show that if baryogenesis occurs prior to dark matter freeze-out (as common in Affleck-Dine models), and the dark matter relic density is diluted by a subsequent entropy dump, then the bound on thermal relic dark matter from perturbative unitarity is relaxed to $m_{\rm DM} \lesssim 10^{10}$ GeV. This relation makes manifest an intriguing connection between high scale baryogenesis and the maximum mass of freeze-out dark matter, assuming the dark matter abundance is diluted by an entropy injection.

Motivated in part by the link between high scale baryogenesis and heavy dark matter, we proceed to study models of ``superheavy asymmetric dark matter,'' in which the dark matter relic density is determined by a particle asymmetry. We show that the presence of a moderate entropy injection, which simultaneously dilutes the dark matter number density, and asymmetries in the baryonic and dark sectors, naturally accommodates superheavy asymmetric dark matter. Intriguingly, it has been argued that the accumulation of asymmetric dark matter 
with mass 0.1-100 PeV in stellar objects can lead to pulsar collapse in the Milky Way galactic center \cite{Bramante:2014zca,Bramante:2015dfa} and ignition of type-Ia supernovae \cite{Bramante:2015cua} (see \cite{Graham:2015apa} for related work), both of which are open problems in astrophysics. 

The paper is structured as follows; we begin in Section \ref{sec:unibar} by deriving the unitarity bound on the dark matter mass for the case of high scale baryogenesis and a period of entropy injection following dark matter freeze-out.  In Section \ref{sec:adm} we consider superheavy asymmetric dark matter models and show that sizable entropy injections which dilute both the frozen out dark matter abundance, and baryon and dark matter asymmetries permits superheavy asymmetric dark matter. Section \ref{sec:3entropy} quantifies the magnitude of entropy dumps from decaying states in the early universe.  Large baryon asymmetries from high scale (Affleck Dine) baryogenesis motivates large entropy dumps and in Section \ref{sec:asymm} we discuss specific implementations within our framework, with focus on generating modest hierarchies between the baryon and dark matter asymmetries.  In Section \ref{sec:5} we present some concluding remarks and comment on possible connections to models of High Scale Supersymmetry.


\section{Dark Matter Mass Upper Bound for Freeze-out after Baryogenesis}
\label{sec:unibar}

While the specific origin of the matter-antimatter asymmetry in the universe is presently unknown, the broad features of primordial asymmetry generation are understood. If a state carries a baryon number $B$, in the presence of out-of-equilibrium effects which violate $B$ and $CP$, an asymmetry can arise such that there is a net number density between the baryons and antibaryons
\begin{align}
\eta_B \equiv n_{\rm B}/s\equiv(n_{\rm b}-n_{\rm \bar{b} })/s,
\label{eq:etadef}
\end{align}
where $s$ is defined as the entropy density of the thermal bath and $n_{\rm b},n_{\rm \bar{b} } $ are the number densities of baryons and antibaryons.  Analogous asymmetries can arise for other global charges, and such asymmetries may also be connected to dark matter \cite{Zurek:2013wia,Petraki:2013wwa}.

It is notable that Affleck-Dine (AD) baryogenesis often leads to particle asymmetries as large as $\eta_B^{\rm initial} \sim \mathcal{O}(1)$, but generally no larger \cite{Linde:1985gh}. Indeed, as discussed in Section \ref{sec:asymm}, large initial asymmetries are the typical expectation. Thus in order for the AD mechanism to yield the observed baryon asymmetry $\eta_B^{\rm now}\sim10^{-10}$, one requires subsequent dilutions by a factor $\zeta \sim \eta_B^{\rm initial}/\eta_B^{\rm now}$. A dilution factor $\zeta$ can arise, for example, if a heavy state decays at late times into the primordial thermal bath
\begin{align}
\zeta\equiv s_{\rm before}/s_{\rm after},
\end{align}
where ``before" and ``after" indicate the entropy density of the thermal bath immediately before and after the decay of the heavy state.
We shall be initially agnostic about the precise source of this entropy injection, simply parameterizing it with $\zeta$, but we will discuss the provenance and magnitude of $\zeta$ in Section \ref{sec:3entropy}.
 
Since the freeze-out abundance $Y\equiv n/s$ depends upon the entropy density $s$ relative to the frozen out number density $n$, a late entropy injection can dilute the dark matter abundance by a potentially large factor.
Crucially, observe that if this dilution occurs after dark matter has frozen out to a fixed abundance in the early universe, then the dark matter abundance will also be diluted by a factor $\zeta$
\beq
\Omega_{\rm DM}^{\rm Relic}  &\simeq \zeta\times \Omega_{\rm DM}^{\rm FO},
\label{eq:omegadil}
\eeq
where the observed value is $\Omega_{\rm DM}^{\rm Relic} h^2\simeq0.12$ \cite{Ade:2015xua}. As we will see, the possibility of late time entropy injection is particularly salient for heavy dark matter. 

For simplicity, we will restrict our attention to the case that dark matter freezes out from a radiation-dominated universe,\footnote{More generally, dark matter may decouple during matter domination. During matter-domination $H\propto T^4$ (rather than $H\propto T^2$) \cite{Giudice:2000ex,Gelmini:2006pw}, which substantially alters the freeze-out calculation.} with an abundance that is later diluted by a factor $\zeta$. The evolutions of particle abundances $Y$ are customarily tracked with respect to the dimensionless temperature variable $x\equiv m_{\rm DM}/T$. Assuming the particles are stable over the lifetime of the universe, these abundances remain constant after particle annihilations cease and the particle has ``frozen-out." For weakly interacting particles, this typically occurs for $x \sim 10$. 
The self-annihilation cross-section of dark matter can be expanded in powers of inverse $x$:  
$\langle\sigma v \rangle \equiv \sum_{n=0} \sigma_n x^{-n} = \sigma_0+\sigma_1 x^{-1}+\OO(x^{-2}),$
 where these give the $s$-wave, $p$-wave, etc. annihilation components.\footnote{If the mediators of the annihilations are light compared to $m_{\rm DM}$ this expansion is not always valid \cite{Griest:1990kh}.}  We can often approximate $\langle\sigma v\rangle$ by the lowest order non-vanishing term in its expansion. The temperature at which dark matter annihilations freeze-out is well described by
\cite{Kolb:1990vq}
\begin{align}
x_{\rm FO} \simeq {\rm ln} \left(K \right) -\left( \frac{1}{2} + n \right) {\rm ln} \left[ {\rm ln} \left(  K  \right) \right], 
\label{eq:SF}
\end{align}
in terms of $K\equiv a(n+1)\sqrt{\frac{\pi}{45}}\sqrt{ g_\star}M_{\rm Pl}m_{\rm DM} \sigma_n$, where $a\simeq 0.145g/g_{\rm \star S}$ for dark matter with $g$ internal degrees of freedom, for a thermal bath with  $g_\star$ massless degrees of freedom, and $g_{\rm \star S}$ entropy-normalized massless degrees of freedom, as defined in \cite{Kolb:1990vq}.  Note, for the Standard Model $g_\star= g_{\rm \star S} \simeq 107$ at temperatures in excess of $200~{\rm GeV}$.

Here we consider the scenario that particle dark matter reproduces the observed dark matter relic abundance through freeze-out to an over-abundance during radiation domination, followed by a period of dilution. The relic density of freeze-out dark matter followed by subsequent entropy injection (cf.~Eq.~\eqref{eq:omegadil}) is
 \beq
\Omega_{\rm DM}^{\rm Relic}  h^2
\simeq
\zeta\times\left[10^9 \frac{\sqrt{g_\star} (n+1) x_{\rm FO}^{n+1}}{g_{\rm \star S} M_{\rm Pl} \sigma_n {\rm GeV}}\right].
\label{eq:simplefreezeout}
\eeq
The numerical prefactors in Eq.~\eqref{eq:simplefreezeout} are for Majorana fermion dark matter, although this can be easily adapted, $e.g.$ for a Dirac fermion, by multiplying by a factor of two. 

\begin{figure}[t!]
 \begin{center}
\includegraphics[width=0.55\textwidth]{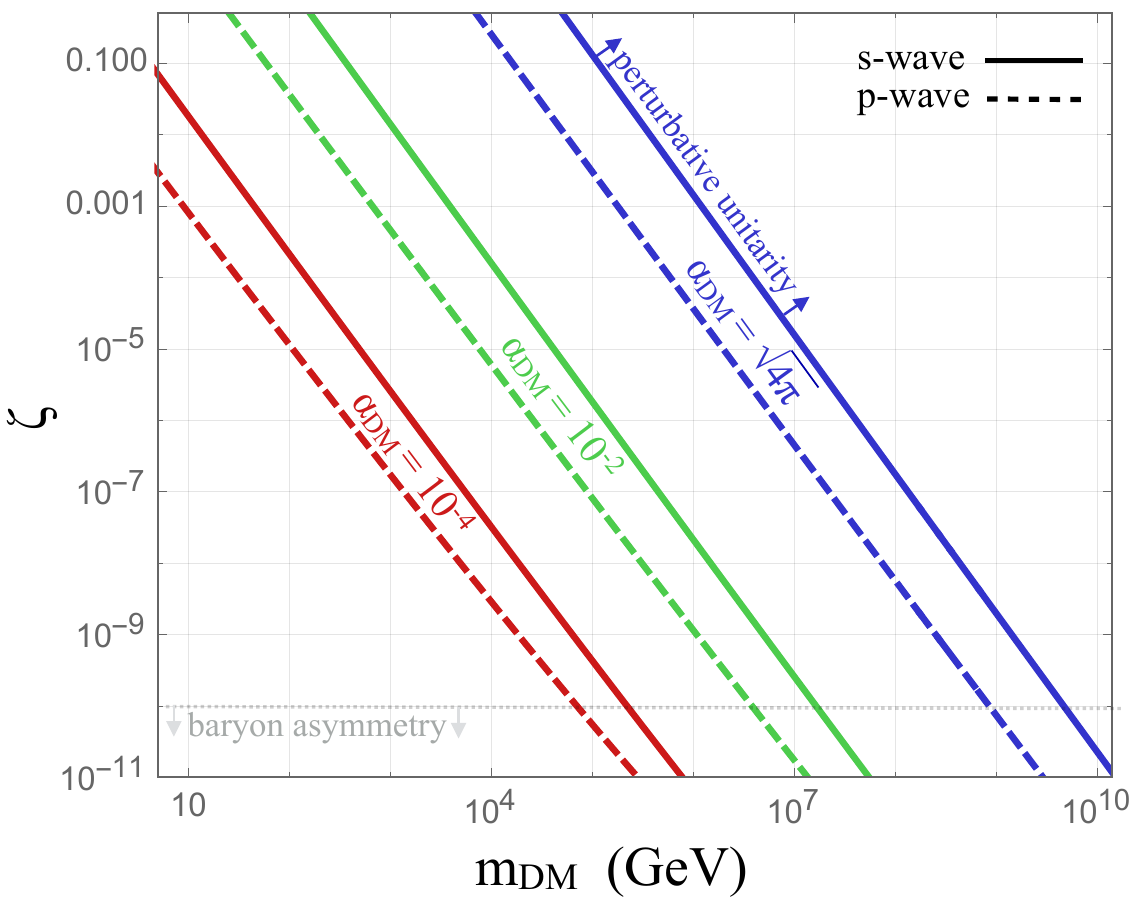}
\end{center}
\vspace{-3mm}
\caption{The late time number density dilution factor $\zeta$ required to match the observed dark matter relic abundance, is plotted against the dark matter mass $m_{\rm DM}$, for indicated choices of $\alpha_{\rm DM}$. Specifically we consider dark matter with $n=0$ $s$-wave (solid)  and $n=1$ $p$-wave (dashed) annihilation cross sections, with $\sigma=\alpha_{\rm DM}^2/m_{\rm DM}^2x^{-n}$. Assuming high scale baryogenesis, the requirement that the baryon asymmetry is not diluted below the observed value imposes $\zeta~ \gtrsim ~10^{-10}$. Further requiring that the dark matter self-annihilation cross-section not exceed the perturbative unitarity bound, $\alpha_{\rm DM} \lesssim \sqrt{4 \pi}$, implies an upper mass bound on thermal dark matter, $m_{\rm DM}\lesssim10^{-10}$ GeV.
\label{fig:unib}}
\end{figure}

Next we specify the dark matter annihilation cross-section $\sigma_n$ in Eq.~\eqref{eq:simplefreezeout} and calculate the dark matter relic density as a function of dark matter mass $m_{\rm DM}$, coupling strength $\alpha_{\rm DM}$, and dilution factor $\zeta$. We take the simplest scenario of dark matter freeze-out via $s$-wave annihilations  ($n=0$), as occurs if the dark matter annihilations to quarks through a vector mediator $V$. Specifically, suppose that the mass of the dark mediator is the same scale as the dark matter, $m_V \sim m_{\rm DM}$ and parameterize the $s$-wave cross section as follows 
 \beq
 \sigma_0 \sim \alpha_{\rm DM}^2 / m_{\rm DM}^2~.
\label{eq:sig0}
 \eeq
In this case the relic dark matter abundance (for $n=0$) is
\beq
\Omega_{\rm DM}^{\rm Relic} h^2 &\simeq 0.1 \left( \frac{m_{\rm DM}}{ {\rm PeV}} \right)^2  \left(\frac{0.3}{\alpha_{\rm DM}}\right)^2 \left(\frac{\zeta}{ 10^{-5}}\right).
\label{eq:value}
\eeq
The size of $\zeta$ required to reproduce the observed relic density is shown in Figure \ref{fig:unib} for $s$-wave and $p$-wave cases.
The dilution factor indicated in Eq.~\eqref{eq:value} of $\zeta\sim10^{-5}$ implies the initial baryon asymmetry is required to be $\eta_B^{\rm initial} \sim 10^{-5}$, since the final baryon asymmetry will be
\beq
\eta_B^{\rm final} = \eta_B^{\rm initial} \zeta \simeq 10^{-10} \left(\frac{\eta_{B}^{\rm initial}}{10^{-5}} \right) \left(\frac{\zeta}{10^{-5}} \right)~.
\label{eq:baryinitial}
\eeq
In the Affleck-Dine scenario it has been argued  \cite{Linde:1985gh} that there is an upper bound on the magnitude of asymmetry that can be generated
\beq
\eta^{\rm initial}_B\lesssim1,
\label{eq:res1}
\eeq
where an $\mathcal{O}(1)$ asymmetry can be generated if a baryon-charged field dominates the energy density of the universe when it decays. We are unaware of well-motivated mechanisms which can yield larger (or even comparable) asymmetries.

Perturbative unitarity \cite{Griest:1989wd} requires that the annihilation cross-section be smaller than
\begin{align}
\sigma_0 \lesssim 4 \pi /m_{\rm DM}^2.
\label{eq:res2}
\end{align}
Using Eq.~\eqref{eq:value} to ensure that the observed relic density is reproduced, and applying restrictions from Eq.~\eqref{eq:res1} and \eqref{eq:res2}, we derive the following upper bound on the mass of thermal dark matter which freezes-out through perturbative $s$-wave annihilations,
\begin{align}
m_{\rm DM}\lesssim 10^{10}~{\rm GeV}.
\label{limit}
\end{align}
This can also be inferred directly from Figure \ref{fig:unib}. It is straightforward to generalize this bound to annihilation cross-sections that are not predominantly $s$-wave  ($n\geq1$).

Because this bound applies to a definite cosmological history (high scale baryogenesis and dark matter freeze-out, followed by dilution), there are a number of caveats, but they do require some model building to realize. Specifically, we can list a number of ways that dark matter could be heavier:
\begin{itemize}
\item Low scale baryogenesis, with $\eta_B$ (re)generated after an entropy injection.
\item Dark matter could freeze-out during a period of matter domination or reheating \cite{Giudice:2000ex,Gelmini:2006pw}. 
\item The dark matter mass could evolve to larger values at late time, after dark matter freeze-out, due to the evolution of a scalar potential that sets its mass \cite{Hui:1998dc}.  
\item The dark matter could form heavy bound states after freeze-out of the Standard Model thermal bath, as in ``Atomic Dark Matter'' \cite{Kaplan:2009de,Hardy:2014mqa}.
\end{itemize}
Even with these provisos, the class of models to which our arguments apply is broad. Indeed, Affleck-Dine baryogenesis and late time entropy production are common features in Standard Model UV completions in SUSY and string theory.

Before moving on it is interesting to note that since the dark matter is overproduced prior to the entropy injection it can have much smaller couplings than un-diluted thermal relic dark matter. From inspection of Eq.~\eqref{eq:simplefreezeout} the annihilation cross section needed to reproduce the relic density, relative to standard freeze-out, is reduced by a factor of $\zeta$. For many models of dark matter, this will relax direct detection constraints whenever $\zeta\ll1$. At a rough, order of magnitude level, if we assume the per-nucleon dark matter direct detection scattering cross-section $\sigma_{\rm N}$, is approximately the size of the dark matter self-annihilation cross-section $\sigma_{\rm N} \sim \sigma_0$, we can surmise that some portions of superheavy dark matter parameter space lie at direct detection cross-sections below the atmospheric and solar neutrino background. 

At high masses ($m_{\rm DM} > 100~{\rm GeV}$), the cross section at which solar and atmospheric neutrinos provide a substantial background to direct detection experiments is
\begin{align}
\sigma^{\rm \nu Floor}_{\rm N}
\sim10^{-20}~{\rm GeV}^{-2}\left(\frac{m_{\rm DM}}{1~{\rm TeV}}\right)~.
\end{align}
The annihilation cross section required to match $\Omega_{\rm DM}^{\rm Relic}  h^2$  is $\sigma_0\simeq
\zeta\times10^{-10} {\rm GeV}^{-2}   $ (taking $x_{\rm FO}\sim20$). Therefore, assuming $\sigma_{\rm N} \sim \sigma_0$, the dark matter direct detection signal lies above the neutrino background whenever
 \beq
\frac{\sigma_0}{\sigma^{\rm \nu Floor}_{\rm N}} \sim100 \times \left(\frac{\zeta}{10^{-5}}\right) \left(\frac{1~{\rm PeV}}{m_{\rm DM}}\right)
\gtrsim 1.
\eeq
The values indicated are chosen to match Eq.~\eqref{eq:value}, thereby demonstrating that superheavy dark matter may be found before solar and atmospheric neutrinos provide a significant background to xenon direct detection experiments. There are some studies of direct detection \cite{Albuquerque:2003ei} and indirect detection \cite{Albuquerque:2000rk,Blasi:2001hr} of non-thermal superheavy dark matter. We leave the investigation of methods for finding superheavy thermal dark matter to future work.


\section{Asymmetric Dark Matter \& Entropy Injection}
\label{sec:adm}

If dark matter carries a global charge, an asymmetry between dark matter and anti-dark matter can arise which is responsible for setting the dark matter relic density  \cite{Zurek:2013wia,Petraki:2013wwa}. Henceforth for concreteness we shall assume that the dark matter is a Dirac fermion (it could equally be a complex scalar). If the dark asymmetry determines the relic abundance, then $\eta_{\rm DM} \equiv  (n_{\rm DM} - n_{\rm \overline{DM}})/s$, defined analogous to Eq.~\eqref{eq:etadef}, directly determines $\Omega_{\rm DM}^{\rm Relic}$ via
\beq
\frac{\Omega_{\mathrm{DM}}^{\rm Relic}}{\Omega_{B}^{\rm Relic}}= \frac{m_{\rm DM}~ \eta_{\rm DM}^{\rm now}}{m_p~\eta_B^{\rm now}}\approx5.5~,
\label{relic}
\eeq
where  $m_p\approx0.94$ GeV is the proton mass, and we note that the observed ratio of dark-to-baryonic matter is approximately $5.5$ \cite{Ade:2015xua}.
For example, normalizing to PeV mass asymmetric dark matter, the final asymmetry needed to match the observed dark matter relic density is
\beq
\frac{\Omega_{\mathrm{DM}}^{\rm Relic}}{\Omega_{B}^{\rm Relic}}\simeq  \left(\frac{m_{\rm DM}}{1~{\rm PeV}}\right)\left(\frac{\eta_{\mathrm{\rm DM}}^{\rm now}}{6 \times 10^{-16}}\right)~.
\label{omega5}
\eeq

For the asymmetry $\eta_{\rm DM}$ to determine the relic density the symmetric component of the dark matter population must annihilate away, so that mostly the asymmetric component remains \cite{Lin:2011gj,MarchRussell:2012hi}. As a result the dark matter mass can typically be constrained by unitarity arguments \cite{Griest:1989wd} to be $m_{\rm DM}\lesssim100$ TeV (assuming dark matter annihilates via perturbative processes).  However, as illustrated in Section \ref{sec:unibar}, entropy injection ($e.g.$ from a late-decaying field) can dilute both symmetric and asymmetric dark matter components, thereby evading the na\"ive unitarity bound.

Hereafter, we will examine a scenario in which the asymmetries $\eta_B$ and $\eta_{\rm DM}$ are too large in the early universe, compared to their values today. As we will see, it is possible for dark matter with PeV-EeV mass to have a perturbative annihilation rate large enough to reduce the symmetric dark matter component below the contribution due to the asymmetry. In this case, both the asymmetric and symmetric dark matter components will be \emph{initially} larger than the observed relic abundance. A subsequent period of entropy production dilutes the symmetric and asymmetric components of the dark sector, along with the baryon asymmetry, altogether yielding the abundances observed today.\footnote{Related ideas on asymmetry dilution have arisen in e.g.~\cite{Randall:2015xza,Davoudiasl:2015vba,Allahverdi:2010rh}.}

The abundance of dark matter prior to the entropy injection, but after it freezes out of the thermal bath, is given by
\beq
\Omega_{\rm DM}^{\mathrm{FO}}h^2
&= \Omega_{\mathrm{Sym}}^{\mathrm{FO}}h^2+ \Omega_{\mathrm{Asym}}^{\mathrm{FO}}h^2~.
\label{om1}
\eeq
The first term of Eq.~\eqref{om1} corresponds to the symmetric abundance of dark matter-anti dark matter pairs, the latter term is the abundance due to the asymmetry.  Following the entropy dump the quantities $Y_{\rm Sym}$ and $\eta_{\rm DM}$ are both reduced by a factor of $\zeta$. Therefore, the present day relic density is altogether given by
\beq
\Omega_{\rm DM}^{\mathrm{Relic}}
&= 
\frac{s_0 m_{\mathrm{DM}}}{\rho_c} \zeta \left[Y_{\mathrm{Sym}}^{\mathrm{FO}}+ \eta_{\mathrm{DM}}^{\mathrm{FO}} \right]~,
\label{eq:finalasym}
\eeq
where $s_0\approx 2.8\times10^3~{\rm cm}^{-3}$ is the entropy density today  and  $\rho_c\approx10^{-5} h^2~{\rm GeV}~{\rm cm}^{-3}$ is the critical density. For the asymmetry to determine the final relic density, inspection of Eq.~\eqref{eq:finalasym} reveals that after freeze-out, the symmetric abundance must satisfy  $Y_{\mathrm{Sym}}^{\rm FO} \ll \eta_{\mathrm{DM}}^{\rm FO} $. This requires that the freeze-out dark matter annihilation cross section is large enough to deplete $Y_{\mathrm{Sym}}$ to a size smaller than $\eta_{\rm DM}$.

 \begin{figure}[t]
\begin{center}
\includegraphics[width=0.55\textwidth]{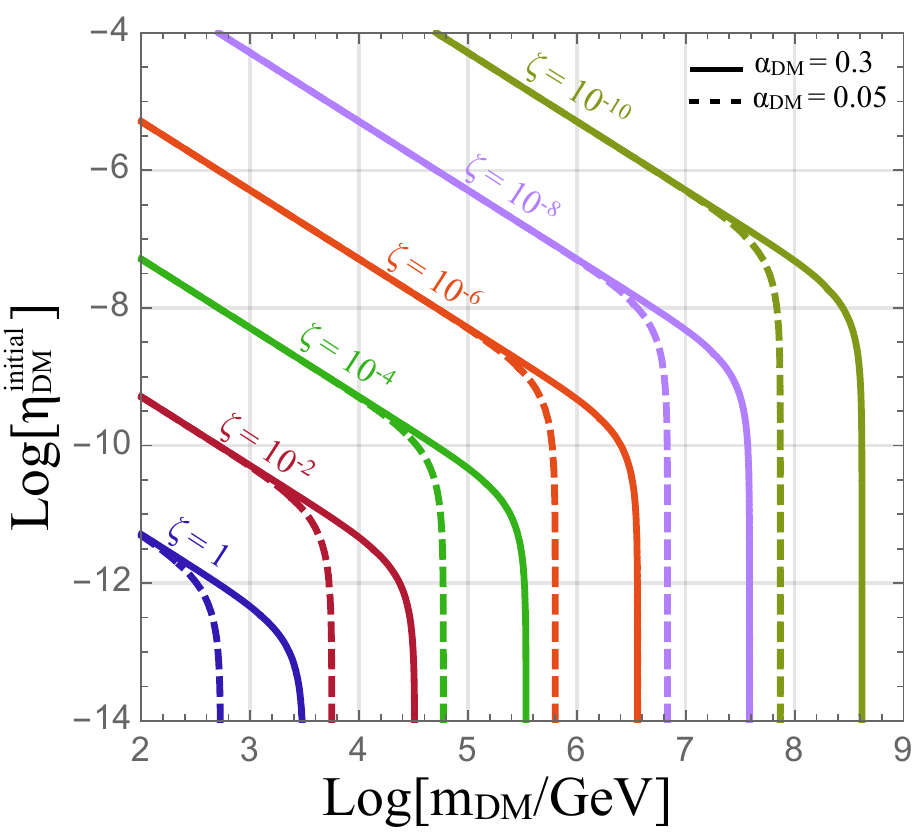}
\end{center}
\vspace{-3mm}
\caption{
Contour plots of the initial dark asymmetry plotted against dark matter mass, with late time dilution factor $\zeta$ as indicated, so that the dark matter fulfils the relic abundance requirement $\Omega_{\rm DM} h^2 \simeq 0.12$. We show this for the case of s-wave ($n=0$) dark matter annihilation with cross-section $\sigma_0 = \alpha_{\rm DM}^2 / m_{\rm DM}^2$, where couplings of $\alpha_{\rm DM} = 0.3$ (solid) and $\alpha_{\rm DM} = 0.05$ (dashed) have been plotted. The contours transition from diagonal to vertical, as the final asymmetric abundance becomes subdominant to the symmetric abundance ($i.e.$ having equal parts particle and antiparticle). Observe that for more weakly coupled dark matter, this transition occurs at lower dark matter masses.
\label{Fig2}}
\end{figure}

 The contribution to the freeze-out abundance from the symmetric component of dark matter after freeze-out, assuming freeze-out from a radiation dominated universe and a subsequent entropy dilution $\zeta$, is given by
\begin{align}
\Omega_{\rm Sym}^{\rm Relic}h^2
=\zeta~\frac{s_0 \rho_{\rm FO}}{\rho_c s_{\rm FO}}~\Omega_{\rm Sym}^{\mathrm{FO}}h^2
\simeq \zeta \times\left[\frac{2\times10^9 \sqrt{g_\star} (n+1) x_{\rm AF}^{n+1}}{g_{\rm \star S}M_{\rm Pl} \sigma_n {\rm GeV}}\right]~.
\end{align} 
The term in brackets is the standard symmetric freeze-out expression for a Dirac fermion (note the extra factor of two compared to \eqref{eq:simplefreezeout}). However, the point of freeze-out  $x_{\rm AF}$ is modified due to the asymmetry, and in the limit $T_{\rm FO} \gtrsim 100 ~{\rm GeV}$ can be approximated as  \cite{Graesser:2011wi} (see also \cite{Iminniyaz:2011yp})
\beq
 x_{\rm AF}\simeq{\rm Log}\left[K\right]
 +\frac{1}{2} {\rm Log} \left[\frac{ {\rm Log}^3\left[K\right]}{ {\rm Log}^{2n+4}\left[K\right]-g_{\star S}(\eta_{\rm DM}^{\rm initial} \frac{K}{2a})^2} \right].
 \label{eq:AF}
\eeq
Freeze-out still typically occurs for $x_{\rm AF}\sim \mathcal{O}(10)$ and remains only logarithmically sensitive to changes in cross-section and mass.  Taking a value $x_{\rm AF}\simeq20$ which is characteristic for PeV mass asymmetric dark matter, the relic abundance is parametrically 
\beq
\Omega_{\rm DM}^{\rm Relic} h^2 \simeq 0.01 \left( \frac{m_{\rm DM}}{ {\rm PeV}} \right)^2  \left(\frac{0.3}{\alpha_{\rm DM}}\right)^2 \left(\frac{\zeta}{10^{-6}}\right) 
+ 0.1 \left( \frac{\eta_{\rm DM}^{\rm initial}}{5\times10^{-10}} \right) \left(\frac{m_{\rm DM}}{{ \rm PeV}} \right)  \left(\frac{\zeta}{10^{-6}}\right).
\eeq
Comparing the first term ($\Omega_{\rm Sym}^{\rm relic}$) to the latter ($\Omega_{\rm Asym}^{\rm relic}$), we see that for suitable parameter values, the condition $\Omega_{\rm Sym}^{\rm relic} \ll \Omega_{\rm Asym}^{\rm relic}$ is satisfied. The viable parameter space is illustrated in Figure \ref{Fig2}. As can be seen, in the presence of a sizeable entropy injection after dark matter freeze-out, models of PeV-EeV mass asymmetric dark matter can reproduce the observed dark matter relic abundance, given a suitably large initial dark and baryon asymmetry.

\section{Entropy from Decays}
\label{sec:3entropy}

Thus far we have treated $\zeta$ as a free parameter. In this section we examine mechanisms that lead to entropy injection in order to quantify the magnitude of $\zeta$. We subsequently discuss the model building implications and constraints on such scenarios. In Section \ref{sec:asymm} we will highlight the importance of entropy injection for obtaining the baryon (and DM) asymmetries in Affleck-Dine models.

\subsection{Magnitude of the Entropy Injection}

Entropy injection can come from a variety of sources, perhaps the most typical are heavy states decaying to the thermal bath, $e.g.$~\cite{Dine:1995kz,Randall:2015xza,de Carlos:1993jw,Banks:1993en,Allahverdi:2010rh,Davoudiasl:2015vba,Berlin:2016vnh}, and phase transitions \cite{Wainwright:2009mq}.  Henceforth, we focus on the former, in which the entropy injection is due to a state $\chi$, which comes to dominate the energy density of the universe after dark matter freezes-out, and subsequently decays to Standard Model states. 

In order for a substantial dilution to take place, we require that the energy density in $\chi$ when it decays greatly exceeds the energy density in all other fields in the universe. The entropy jump in the Standard Model radiation bath due to the decays of $\chi$ is given by
\beq
\zeta\equiv \frac{s_{\rm before}}{s_{\rm after}}\simeq\left(\frac{\rho_{\rm rad}}{\rho_\chi}\Big|_{H=\Gamma_\chi}\right)^{3/4}~,
\label{zeta}
\eeq 
where $\rho_{\rm rad,\chi}$ is the energy density in radiation bath and $\chi$ states respectively.
At the time of decay ($H \sim \Gamma_\chi$) the energy density in $\chi$ is $\rho_\chi=\sqrt{3 \Gamma_\chi M_{\rm Pl}}$.  Below some critical temperature, the energy density of $\chi$ starts evolving as $a^{-3}$ (matter-like), compared to the radiation bath which redshifts like $a^{-4}$ (radiation-like), where here $a$ is the standard FRW scale factor. This relative evolution leads to $\chi$ coming to dominate the energy density of the universe. 

There are primarily two reasons for $\rho_\chi$ to have matter-like evolution in the early universe: (i.) $\chi$ is a particle that is non-relativistic and thermally decoupled from the rest of the universe; (ii.) $\chi$ is a light, slowly decaying bosonic field oscillating in its potential, so that its average equation of state is $w \sim 0$ ($i.e.$ matter-like). In the first case (i.) $\chi$ starts evolving as $a^{-3}$ at $T_{\rm crit}\sim m_\chi$, when the temperature of the thermal bath drops below $m_\chi$ and its momentum becomes negligible. For case (ii.) $\chi$ becomes matter-like when the $\chi$ field begins to oscillate around its minimum when $m_\chi \sim H$ or equivalently $T_{\rm crit}\sim \sqrt{3 m_{\chi} M_{\rm Pl}}$, assuming a simple quadratic potential for $\chi$, $i.e.$ $V \supset m_\chi^2 \chi^2$.

We will restrict our attention to models in which dark matter freeze-out occurs prior to the energy density of $\chi$ becoming matter-like, $T_{\rm FO}\sim m_{\rm DM}\gg T_{\rm crit}$. In this case freeze-out occurs during radiation domination. When $\chi$ decays it reheats the thermal bath to a temperature $T_{\rm RH}$ and dilutes asymmetries and frozen out abundances. For $m_{\rm DM}>m_\chi$ decays of $\chi$ to dark matter are kinematically forbidden, thus the dark matter is diluted and not repopulated during $\chi$ decays, and for $T_{\rm FO}\gg T_{\rm RH}$, interactions in the thermal bath will no longer produce dark matter states.

The Friedman equation giving the evolution of the energy density for $H(T_{\rm crit})>H>\Gamma_{\chi}$ is given by
\begin{align}
H^2\simeq \frac{\pi^2}{90}\frac{ g_\star T_{\rm crit}^4}{M_{\rm Pl}^2}\left[R_\chi\left(\frac{1}{\Delta a}\right)^{3}
+R_{\rm rad}\left(\frac{1}{\Delta a}\right)^{4}\right]~,
\label{eq:FE}
\end{align} 
where $\Delta a\equiv a(T)/ a(T_{\rm crit})$ is the change in the scale factor after $\rho_\chi$ became matter-like, and $R_i\equiv \rho_i/(\rho_\chi+\rho_{\rm rad})\big|_{\rm crit}$ is the relative energy densities of $\chi$ and the Standard Model radiation at some initial point in time, in this case at the time when $T=T_{\rm crit}$. As one example, note that if $\chi$ is a particle initially in thermal equilibrium with the radiation bath, but has an extremely weak self-annihilation cross-section, then $R_\chi \simeq R_{\rm rad}g/g_\star \sim R_{\rm rad}/100$ \cite{Kolb:1990vq}. Conversely, if $\chi$ is a scalar field oscillating in its potential, or was produced by an out-of-equilibrium decay, then potentially $R_\chi/ R_{\rm rad}\simeq1$. 
Note that in Eq.~\eqref{eq:FE} we have neglected the contribution from dark matter, since this is Boltzmann suppressed after freeze-out $\rho_{\rm DM}\propto {\rm exp}(-x_{\rm FO})\ll1$, and for the cosmological epochs we are considering, will not come to dominate the energy density of the universe.

For $H(T_{\rm crit})>H>\Gamma_{\chi}$ the contribution from $\chi$ grows and becomes comparable to the radiation energy density at $T=T_{\rm MD}$, or after a period
\beq
\Delta a_{\rm MD}\equiv\frac{a(T_{\rm MD})}{a(T_{\rm crit})}\simeq\frac{R_{\rm rad}}{R_\chi}~.
\label{eq:FE2}
\eeq
The $\chi$ energy density continues to grow until it decays to radiation at $H \sim \Gamma_{\chi}$, this occurs after 
\beq
\Delta a_\Gamma\equiv \frac{a(H=\Gamma_{\chi})}{a(T_{\rm crit})}\simeq
\left(\frac{\pi^2 g_\star T_{\rm crit}^4}{90 M_{\rm Pl}^2\Gamma_{\chi}^2}R_\chi\right)^{1/3}~,
\label{eq:deltaa}
\eeq
where in deriving Eq.~\eqref{eq:FE2} and \eqref{eq:deltaa} we have assumed that $\chi$ is sufficiently long lived that it dominates  Eq.~\eqref{eq:FE}, otherwise the entropy change would be negligible. 
We can find $\rho_\chi$ at the time of $\chi$ decay $H \sim \Gamma_{\chi}$ by evolving $\chi$'s energy density with Eq.~(\ref{eq:deltaa}) to obtain
\begin{align}
\rho_{\chi}\big|_{\Gamma_\chi} &=\rho_{\chi}\big|_{T_{\rm crit}}\Delta a_{\Gamma}^{-3} \nonumber \\
&=\frac{g_\star \pi^2}{30} T_{\rm crit}^4 R_\chi  \Delta a_{\Gamma}^{-3}\simeq 3 \Gamma_{\chi}^2M_{\rm Pl}^2~.
\label{eq:rhoxevo}
\end{align}
We can also find $\rho_\chi$ at the time of $\chi$ decay as a function of the reheat temperature
\beq
\rho_\chi\big|_{\Gamma_\chi} \equiv \frac{\pi^2 g_\star(T_{\rm RH})}{30} T_{\rm RH}^4~.
\eeq
Note that in the Standard Model $g_\star(T) \pi^2/30 \simeq 35$ for $T > 200~{\rm GeV}$ \cite{Kolb:1990vq}. On the other hand, the energy density in the radiation $\rho_{\rm rad}$ immediately prior to $\chi$ decay is
\beq
\rho_{\rm rad}\big|_{T_{\rm crit}}\Delta a_{\Gamma}^{-4}
&\simeq
3 \frac{R_{\rm rad} }{R_\chi} \Gamma_{\chi}^2M_{\rm Pl}^2\Delta a_{\Gamma}^{-1}~.
\label{eq:rhorevo}
\eeq
Inserting Eqs.~\eqref{eq:deltaa}-\eqref{eq:rhorevo} into Eq.~\eqref{zeta}, it follows that
\beq
\zeta
\simeq\left(\frac{R_{\rm rad} }{R_\chi}\Delta a_{\Gamma}^{-1}\right)^{3/4}
\sim\left(\frac{R_{\rm rad}^{3/4} }{R_\chi}\right)
\left(\frac{T_{\rm RH}}{T_{\rm crit}}\right)~.
\label{zetader}
\eeq
Assuming that the ratio of energy densities at $T=T_{\rm crit}$ are $R_{\rm rad}/ R_\chi\simeq 1$, the dilution is
\begin{align}
\zeta \sim 10^{-10} \left(\frac{T_{\rm RH}}{10~{\rm MeV}} \right) \left( \frac{10^{8}~{\rm GeV}}{T_{\rm crit}} \right),
\label{eq:zetatemp}
\end{align}
where we normalize to the maximum dilution permitted by high scale baryogenesis, and the reheat temperature after $\chi$ decays $T_{\rm RH} \simeq 10 ~{\rm MeV}$, which is the minimum temperature the Standard Model thermal bath must return to in order to reproduce big bang nucleosynthesis (BBN) observations. 

For the case that dark matter freezes out through s-wave annihilations with cross section $\sigma_0 \sim \alpha_{\rm DM}^2/m_{\rm DM}^2$, Eq.~\eqref{eq:value} combined with Eq.~\eqref{eq:zetatemp} (which assumes $R_{\rm rad}/ R_\chi\simeq 1$) determine the dark matter mass required to match the observed relic for given values of the reheat and critical temperatures 
\begin{align}
m_{\rm DM} \sim 10^9~{\rm GeV} \left(\frac{\alpha_{\rm DM}}{1} \right) \left(\frac{10~{\rm MeV}}{T_{\rm RH}} \right)^{1/2} \left(\frac{T_{\rm crit}}{10^{8}~{\rm GeV}} \right)^{1/2} ~.
\label{eq:mdmfixed}
\end{align}

\subsection{The Dilution Parameter Space}
\label{sec:3B}
The critical temperature at which the evolution of $\rho_\chi$ becomes matter-like is not a free parameter, but is fixed by the details of the model. Below we look at the constraints on $T_{\rm crit}$ corresponding to a decaying state at one time in thermal equilibrium with the radiation bath, $T_{\rm crit}\sim m_{\chi}$, and an oscillating field where $T_{\rm crit}\sim \sqrt{3 M_{\rm Pl}m_{\chi}}$.
For the models outlined in Sections \ref{sec:unibar} \& \ref{sec:adm} to be consistent they are required to satisfy the following criteria:
\begin{itemize}
\item[\em a).] Standard Model reheating (decay of $\chi$) occurs above BBN temperatures.
\item[\em b).] The universe is radiation-dominated during dark matter freeze-out.
\item[\em c).] The entropy jump occurs after freeze-out.
\item[\em d).] $\chi$ dominates the energy density of the universe when it decays.
\end{itemize}
Below we discuss how each of these requirements restricts the parameter space:

\begin{itemize}
\item[\em a).] The Standard Model is reheated above the BBN threshold:~$T_{\rm RH}\simeq\sqrt{\Gamma_{\chi} M_{\rm Pl}}\gtrsim 10$ MeV. From Eq.~\eqref{eq:mdmfixed}, which assumes $R_\chi/R_{\rm rad} = 1$ and a freeze-out annihilation cross-section $\sigma_0 \sim 1/m_{\rm DM}^2 $, it follows that
\begin{align}
T_{\rm crit} \leq 10^{-8} ~{\rm GeV} \left( \frac{m_{\rm DM}}{{\rm GeV}} \right)^2 \left(\frac{1}{\alpha_{\rm DM}^2} \right)^2 \left(\frac{T_{\rm RH}}{10~{\rm MeV}} \right)~.
\end{align}

\item[\em b).] For freeze-out to occur during radiation domination, it is required that $ T_{\rm crit} < T_{\rm FO}$ or
\begin{align}
m_{\chi}\lesssim\left\{
\begin{array}{ccc}
10^9~{\rm GeV}~
\left(\frac{10}{x_{\rm FO}}\right) 
\left(\frac{m_{\rm DM}}{10^{10}~{\rm GeV}}\right)  &~~~~ T_{\rm crit}\sim m_\chi~~~ &({\rm \chi- thermal~particle}) \\[5pt]
1~{\rm GeV}~\left(\frac{m_{\rm DM}}{10^{10}~{\rm GeV}}\right)^2\left(\frac{10}{x_{\rm FO}}\right)^2    & ~~~~T_{\rm crit}\sim \sqrt{3 m_\chi M_{\rm Pl}}~~~ &({\rm \chi -oscillating ~field})
\end{array}\right.~.
\end{align}
Dark matter freeze-out during a period of matter-domination is certainly possible, but the relic density calculation is altered since the Hubble rate is different and the dark matter abundance becomes sensitive to the decay widths of the late decaying scalar $\chi$. Particularly, the decay width of $\chi$ to dark matter can be responsible for setting the dark matter relic abundance \cite{Giudice:2000ex}. We leave a detailed study of superheavy dark matter produced via matter-dominated freeze-out to future work.

\item[\em c).] For the dark matter to be diluted, rather than repopulated, by $\chi$ decays, the lifetime of $\chi$ should be such that $\chi$ decays after dark matter freeze-out. Thus $H(T_{\rm FO})>\Gamma_{\chi}$, or in terms of temperature thresholds $T_{\rm FO} \gtrsim T_{\rm RH} \sim \sqrt{\Gamma_\chi M_{\rm Pl}}$, this implies
\begin{align}
\Gamma_{\chi}\lesssim 10^{-8}~{\rm GeV}\left(\frac{m_{\rm DM}}{1~{\rm PeV}}\right)^2 
\left(\frac{10}{x_{\rm FO}}\right)^2 ~,
\label{eq:c}
\end{align}
or equivalently,
\begin{align}
T_{\rm RH}\lesssim
10^{5}~{\rm GeV}\left(\frac{m_{\rm DM}}{1~{\rm PeV}}\right) 
\left(\frac{10}{x_{\rm FO}}\right) ~.
\label{Trh}
\end{align}
Moreover, using Eq.~\eqref{eq:mdmfixed} this can be expressed in terms of a bound on $T_{\rm crit}$
\begin{align}
T_{\rm crit} \lesssim  10^{-9} ~{\rm GeV} \left( \frac{m_{\rm DM}}{\rm GeV} \right)^3 \left( \frac{1}{\alpha_{\rm DM}} \right)^2 \left( \frac{10}{x_{\rm FO}}   \right)~. 
\end{align}

\item[\em d).] The dark matter energy density should not grow larger than the $\chi$ contribution at any stage after freeze-out, or radiation domination will not be restored after $\chi$ decay. This condition is satisfied for 
\beq
\frac{R_\chi} {R_{\rm rad}}m_{\rm DM} x_{\rm FO}^{3/2}e^{-x_{\rm FO}}<T_{\rm crit}~.
\eeq
For scenarios we have considered, this requirement is redundant when compared to condition {\em (a)}.

\end{itemize}

\begin{figure*}[t!]
\includegraphics[width=0.5\textwidth]{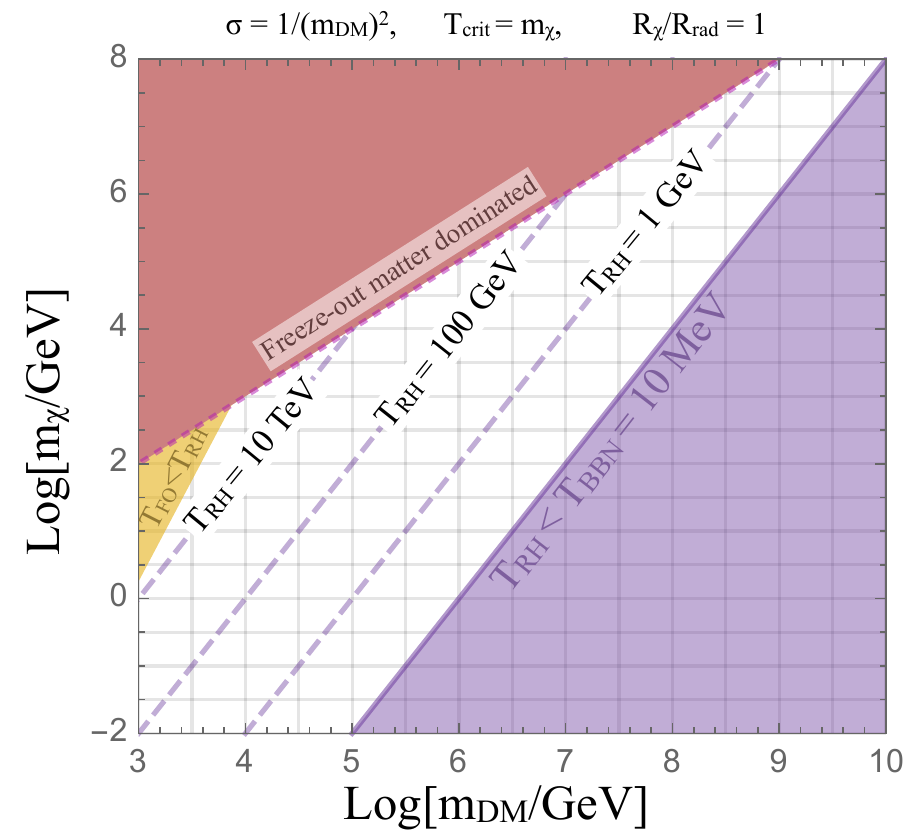}
\includegraphics[width=0.5\textwidth]{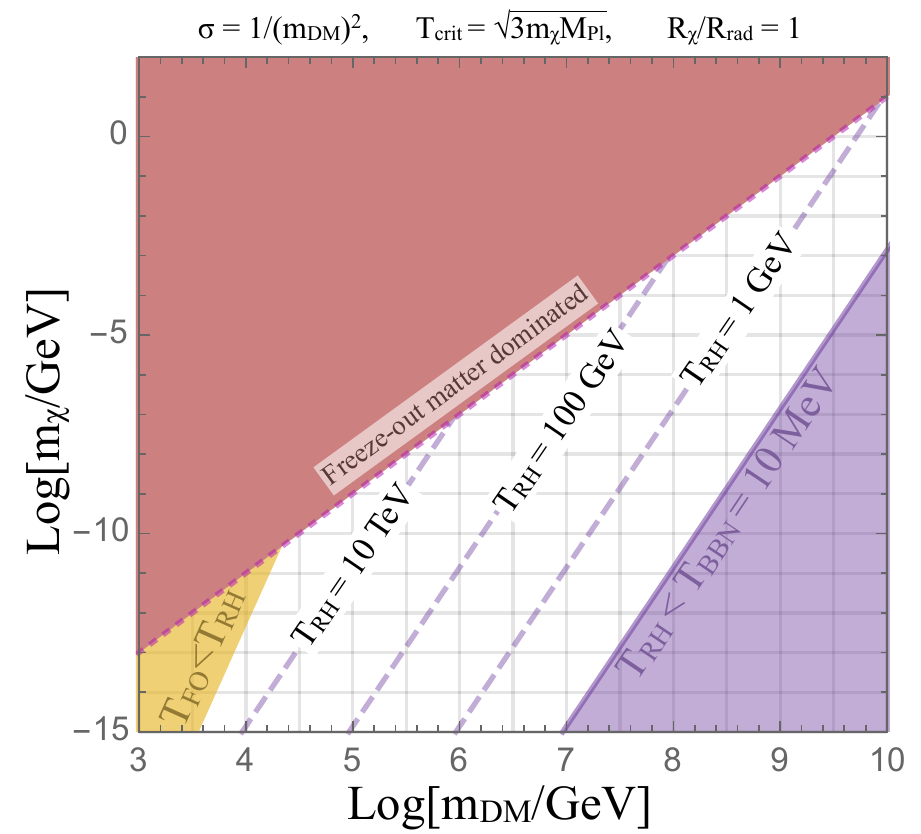}
\caption{
Contours showing the reheat temperature at which $\chi$ decays required to reproduce the observed dark matter relic abundance as a function of $m_\chi$ and $m_{\rm DM}$. These plots assume the energy density in $\chi$ begins matter-like evolution at $T_{\rm crit}\simeq m_\chi$ (left) and $T_{\rm crit}\simeq \sqrt{3 m_\chi M_{\rm Pl}}$ (right). The initial distribution of energy densities in $\chi$ versus the Standard Model thermal bath are $R_\chi/R_{\rm rad} =1$.  The dark matter annihilation cross-section is assumed to be $s$-wave with cross-section $\sigma_0=1/m_{\rm DM}^2$. Shaded regions indicate constraints on the parameter space, as discussed in points (a.)-(c.) in Section \ref{sec:3B}. These requirements substantially restrict the parameter space, but allow for the dark matter mass $m_{\rm DM}$ to be as large as $10^{10}$ GeV.
\label{Fig3}}
\end{figure*}

 Figure \ref{Fig3}  illustrates how these requirements are complementary in constraining the parameter space for both classes of models. It is evident that a range of dark matter and $\chi$ masses reproduce the observed dark matter relic density, while reheating the universe above BBN temperatures. It is interesting to observe that in the parameter space plotted, there is an effective upper bound on the reheat temperature $T_{\rm RH}\lesssim1$ TeV. Higher reheat temperatures imply either freeze-out occurs during matter domination (which changes the freeze-out calculation), or that the dark matter states are repopulated (rather than diluted) following the decays of $\chi$. Notably, for $\chi$ as either a particle or an oscillating field, the dark matter mass is permitted to saturate the upper mass bound of $10^{10}$ GeV, derived in Eq.~\eqref{limit}. 

\section{Affleck-Dine, Dark Matter, and Large Asymmetries}
\label{sec:asymm}

Observing that SUSY models generically present exactly flat directions in the scalar potential in the limit of unbroken SUSY, Affleck and Dine \cite{Affleck:1984fy}  argued that in the early universe it is natural for scalar fields along these flat directions to initially take large field values.\footnote{Analogous mechanisms have been proposed without SUSY. In principal, this can be implemented with a complex scalar that has a global charge and a flat potential (possibly the inflaton), see e.g.~\cite{Linde:1985gh,Delepine:2006rn,Hertzberg:2013mba}.} 
Of primary interest are flat directions which carry a global charge (baryon, lepton, or dark). One can parameterize such flat directions (a product of superfields) in terms of a new superfield, the scalar component of which is commonly dubbed the AD field ($\phi$). Affleck and Dine demonstrated that the evolution of an AD field from its initial field value can generate particles asymmetries, provided that the potential of the AD field violates $C$ and $CP$.  The AD mechanism has since been thoroughly studied  \cite{Dine:2003ax}, including its application to dark asymmetries  \cite{Cheung:2011if,Bell:2011tn}.

In what follows we will examine a minimal AD potential and calculate the resulting particle asymmetry. Our aim is to clarify which models typically lead to large asymmetries $\mathcal{O} (10^{-8})\lesssim \eta_B \lesssim \mathcal{O} (1)$ due to AD baryogensis and thus require significant late time entropy dilution to reproduce the observed level $\eta_B^{\rm now} \sim 10^{-10}$. We will also outline AD dark/co-genesis scenarios with $\eta_B \gg \eta_{\rm DM}$, which is one requirement of the superheavy asymmetric dark matter studied in Section \ref{sec:3entropy}. 
Broadly following \cite{Dine:2003ax}, we take the following AD potential\footnote{It has been noted that a simplified version of this potential suffices for baryogenesis \cite{Cheung:2011if}. Here we retain all the usual terms of the AD potential, so that our treatment of AD dark/co-genesis can be easily ported to full supersymmetric models in future work.} for the complex scalar (AD) field $\phi$
\beq
V_{\rm AD} 
&=  m_\phi^2 |\phi|^2 -   H^2 |\phi|^2 + \frac{|\phi|^{2n+4}}{M^{2n}}  +\left(a H + b m_\phi \right) \left( \frac{\phi^{n+3}}{M^n} + {\rm h.c.}\right) ,
\label{eq:adpot}
\eeq
where $a$ and $b$ are complex numbers, $m_\phi$ is the low-temperature mass of $\phi$ and $M$ is a mass scale at which the higher dimension operator is induced. The potential is comprised as follows:
\begin{itemize}
\item The first term is generated by SUSY-breaking, and becomes relevant for $H \ll m_\phi$.

\item The second, fourth, and fifth terms are generated by inflaton-induced SUSY-breaking. In particular, the last two terms violate baryon number, as required for baryogenesis. In the context of SUSY, the form of these terms arises from inflaton $F$-terms \cite{Dine:2003ax,Dine:1995kz}.

\item  The third term arises from UV corrections at mass scale $M$. The non-renormalisable term
 with the highest power of $\phi$  `lifts' the flat direction when $H \gg m_\phi$, determining the initial minimum of the AD potential.
  \end{itemize}
  
In the early universe, while $H \gg m_\phi$, the AD potential depends mostly on the second and third terms of Eq.~\eqref{eq:adpot}, and has a minimum at
\begin{align}
\phi_0 \simeq  \left(H M^n \right)^{\frac{1}{n+1}}.
\end{align}
As the universe cools, eventually $H\sim m_\phi $, at which point $\phi$ will roll from $\phi_0$ to the new minimum of its potential and undergo coherent oscillations. The baryon (or other charge) asymmetry that arises depends on $\phi_0$, and the relative phase between the couplings $a$ and $b$, which together control the magnitude of $CP$ violation.

Using the equations of motion (the Friedmann equations) for a scalar field in de Sitter space, the change in baryon number is given by \cite{Dine:2003ax,Cheung:2011if}
\begin{align}
\frac{{\rm d} n_B}{{\rm d} t} \simeq \frac{{\rm d} V_{\rm AD}}{{\rm d} \theta}~,
\end{align}
where $\theta$ parameterizes the phase of complex terms carrying AD charge in Eq.~\eqref{eq:adpot}. In this case by construction, the relevant $CP$ and $B$ violating terms are those with coefficients $a$ and $b$. The relative phase of these terms will determine the net baryon charge produced. Although for ${\rm Arg}[a/b]=0$ there will be no net baryon number generated, one may reasonably expect the initial phases to be chosen at random, so that typically ${\rm Arg}[a/b]\sim \mathcal{O}(1)$. With this in mind, henceforth we drop factors of ${\rm Arg}[a/b]$.  

The $\chi$ field starts oscillating when $H\sim m_\phi$ and one can use the approximation\footnote{This approximation has been checked numerically, and tends to agree to within an order of magnitude \cite{Dine:1995kz}.} that $1/t \sim H \sim m_\phi$. It follows that when $\phi$ begins oscillating, the final two terms in Eq.~\eqref{eq:adpot} determine the net charge density created in the universe
\begin{align}
n_{\rm B} &\sim \frac{\phi_0^{n+3}}{M^{n}} \sim  m_\phi^{\frac{n+3}{n+1}} M^{\frac{2n}{n+1}}~.
\label{eq:nbare}
\end{align}
Equation \eqref{eq:nbare} gives the net charge density introduced by the AD field when it begins oscillating, however, the resulting particle asymmetry $\eta_B \sim n_B / s$ also depends upon the relative abundance of other fields in the universe which contribute to $s$, the total entropy density of the universe. We first consider the simplest scenario, that the universe is radiation-dominated when the AD field rolls down its potential and decays. In this case, the asymmetry is given by \cite{Dine:2003ax,Dine:1995kz}
\begin{align}
\eta_B^{(\rm rad)} \simeq \frac{n_{\rm B}}{\rho_{\rm u}^{3/4}}  \simeq \frac{ m_\phi^{\frac{3-n}{2n + 2}}  M^{\frac{2n}{n+1}}}{(3 M_{\rm Pl}^2)^{3/4}}~,
\label{eq:etab}
\end{align}
where we use that the energy density of the radiation-dominated universe is $\rho_{\rm u} \sim 3 m_\phi^2 M_{\rm Pl}^2$ when $\phi$ begins oscillating, which follows from the relationship $H=m_\phi$ and the Friedmann equation $3H^2 = \rho/ M_{\rm Pl}^2$. 

Let us consider some examples cases, which illustrate that the initial asymmetry is often too large. Consider the case where the high-dimension effective operator in Eq.~\eqref{eq:adpot} is mass dimension six ($n=1$), then the resulting asymmetry is
\begin{align}
\eta_B^{(\rm rad, n=1)} \simeq 10^{-8} ~ \left( \frac{m_\phi}{\rm TeV} \right)^{1/2} \left( \frac{M}{M_{\rm Pl}} \right).
\end{align}
We highlight the case of $m_\phi \sim {\rm TeV}$ since it is a typical choice for an AD soft mass term, assuming electroweak-scale supersymmetry.  Furthermore, let us next examine the expected magnitude of asymmetries arising from a higher dimension operator with $n=2$. This implies an even larger initial asymmetry
\begin{align}
\eta_B^{(\rm rad, n=2)} \simeq 10^{-3} ~ \left( \frac{m_\phi}{\rm TeV} \right)^{1/6} \left( \frac{M}{M_{\rm Pl}} \right)^{4/3}.
\end{align}

Such large initial asymmetries require a subsequent dilution mechanism. This problem is even more apparent in the case that the universe is not radiation dominated, but dominated by the energy in the AD  field. If the AD field has an extended period oscillating around its minimum, during which time it redshifts like non-relativistic matter, then its energy density will come to dominate the universe. In this case, the initial particle asymmetry in the universe will be $\eta_{\rm B} \sim 1$ (see \cite{Linde:1985gh} for an extended discussion of this point), in which case a very large late time entropy injection is necessary to match observations.

One way that the required dilution of baryon number is often achieved in studies of AD baryogenesis \cite{Affleck:1984fy,Dine:1995kz,Dine:2003ax}, is by assuming that the inflaton dominates the energy density of the universe, and decays later than the AD field. In this scenario the entropy injection of the decaying inflaton field dilutes the AD asymmetry. However, even in this case, the resulting charge asymmetry can still be much larger than that observed: $\eta_B\gg\eta_B^{\rm now}\simeq10^{-10}$. For $\rho_\phi \ll \rho_{\rm I} $, where $\rho_{\rm I}$ is the energy density of the inflaton field, which is assumed to be oscillating in its potential (diluting like matter $\rho_I \propto a^{-3}$), the asymmetry is
\begin{align}
\eta_B^{\rm (inf)} &\simeq \frac{n_{\rm B}}{\rho_{\rm I}/T_{\rm R,I}} \sim \frac{T_{\rm R,I} m_\phi^{\frac{1-n}{n+1}} M^{\frac{2n}{n+1}}}  {3 M_{\rm Pl}^2}~,
\label{eq:etabinf}
\end{align}
where $T_{\rm R,I}$ is the temperature at which the inflaton field will decay and here we have used $\rho_{\rm I} \sim 3 m_\phi^2 M_{\rm Pl}^2$ at the time that $\phi$ begins oscillating. Specifically, for $n=1$, such that $|\phi|^6$ is the highest dimension operator in the potential of Eq.~\eqref{eq:adpot}, one has
\begin{align}
\eta_B^{\rm (inf,n=1)} \simeq 10^{-10} ~\left(\frac{T_{\rm R,I}}{10^{9}~{\rm GeV}}\right)\left( \frac{M}{M_{\rm Pl}}\right).
\end{align}
This is the standard result in the literature that achieves the observed particle-antiparticle asymmetry using dilution via the late-decay of the inflaton \cite{Dine:1995kz,Dine:2003ax}. However, if the AD mechanism arises from a higher dimension operator ($n=2$), the resulting asymmetry will again be typically too large, even allowing for dilution via subsequent inflaton decay at $T_{\rm R,I} = 10^9 {\rm ~GeV}$, as can be seen from the following expression
\begin{align}
\eta_B^{\rm (inf,n=2)} \simeq 10^{-5} ~\left( \frac{m_\phi}{10^{3}~{\rm GeV}} \right)^{-\frac{1}{3}}  \left( \frac{T_{\rm R,I}}{10^{9}~{\rm GeV}} \right) \left( \frac{M}{M_{\rm Pl}} \right)^{\frac{4}{3}}.
\end{align}
This can be alleviated through stronger dilution due to the inflaton decaying at lower temperatures. Although this approach will run into conflicts with observations if $T_{\rm R,I}\lesssim10$ MeV. Conversely, as has been focus of this paper, an alternative to demanding inflaton energy domination, an entropy injection from a late decaying field can also provide the required dilution of baryon number. 

As detailed in Section \ref{sec:adm}, models of superheavy asymmetric dark matter require the dark sector to have a much smaller matter-antimatter asymmetry than the baryonic sector. As we now show, a large ratio of dark-to-baryon asymmetries, $\eta_B/ \eta_{\rm DM} \gg 1$, can arise if the baryon asymmetry is generated from a higher-dimension operator than the dark asymmetry. 
For simplicity, we assume that the AD field oscillates and decays to Standard Model fields during a radiation-dominated epoch, and is later diluted by a factor of $\zeta$. We make the reasonable simplifying assumption that both the Standard Model and dark AD fields, $\phi_B$ and $\phi_{\rm D}$, have symmetries broken at the same high scale $M$. Then if $\phi_B$ and $\phi_{\rm D}$ have Standard Model and dark asymmetries broken by operators with mass dimension $(4+2j)$ and $(4+2k)$, respectively (cf.~Eq.~\eqref{eq:etab}), then the relative size of the Standard Model and dark asymmetries is given by
\beq
\frac{\eta_{\rm B}^{ (j)}}{\eta_{\rm DM}^{(k)}} \simeq 
 M^{\frac{2j-2k}{(j+1)(k+1)}}\left(\frac{m_{\phi_B}^{\frac{1-j}{j+1}}}{m_{\phi_D}^{\frac{1-k}{k+1}} }\right)~.
\eeq
Note that $j$ or $k=1$ are special since in these cases the ratio is insensitive to $m_{\phi_B}$ or $m_{\phi_D}$, respectively. One might reasonably expect the masses of $\phi_B$ or $\phi_D$ to be comparable since they likely both arise from the same source  of SUSY breaking.

For example, consider the case that the Standard Model asymmetry arises from an operator of leading dimension-6 operator ($j=1$), while the dark asymmetry comes from a leading dimension-8 operator ($k=2$), then the resulting relative asymmetry is
\beq
\frac{\eta_{\rm B}^{ (1)}}{\eta_{\rm DM}^{(2)}} 
&\simeq 
\left(\frac{ M}{m_{\phi_B} }\right)^{\frac{1}{3}}
\simeq 
10^{4}
\left(\frac{ M}{M_{\rm Pl} }\right)^{\frac{1}{3}}
\left(\frac{ 1~{\rm PeV}}{m_{\phi_B} }\right)^{\frac{1}{3}}~.
\eeq
The indicated parameter values are chosen to match the well motivated scenario in which the non-renormalisable operators are generated at the Planck scale, thus $M=M_{\rm Pl}$, and where we have shown $m_{\phi_B} \sim 1~{\rm PeV}$. In this case the expected ratio of the initial asymmetries is $\eta_{\rm B}/\eta_{\rm DM}  \sim 10^4$, which is well suited for the models of superheavy asymmetric dark matter outlined in Section \ref{sec:adm}.


\section{Concluding Remarks}
\label{sec:5}

Traditional models of superheavy dark matter set the observed relic abundance via non-thermal mechanisms such as inflationary dynamics \cite{Chung:1998zb}, gravitational production \cite{Chung:2001cb}, or  thermal inflation \cite{Hui:1998dc}. The scenario we outline here is distinct in that the dark matter undergoes a standard freeze-out process and its abundance is subsequently diluted due to late time entropy production. We have called this scenario ``Superheavy Thermal Dark Matter.'' Moreover, we believe this is the first paper to construct viable models of superheavy \emph{asymmetric} dark matter. 

Thus far we have not specified any UV-completion of superheavy dark matter, but given the links we have drawn to Affleck-Dine baryogenesis, it is interesting to ask whether superheavy dark matter could be the lightest supersymmetric particle of a High Scale SUSY spectrum \cite{Hall:2009nd,Wells:2004di} and thus stable due to $R$-parity. Such High Scale SUSY have been independently motivated via anthropic arguments involving the Higgs mass \cite{Agrawal:1997gf} and provide an interesting alternative to Weak Scale SUSY. Moreover, to realise superheavy SUSY asymmetric dark matter  there are several potential candidates, most prominently Sneutrinos \cite{Hooper:2004dc}, Higgsinos \cite{Blum:2012nf}, or bound states in the hidden sector involved in SUSY breaking \cite{Banks:2005hc,Benakli:1998ut}.

It is also interesting to note that in certain classes of models the Higgs quartic coupling $\lambda$ is anticipated to vanish at the scale of the SUSY partners $M_{\rm SUSY}$. Models which automatically imply the vanishing of the quartic coupling at the SUSY scale occurs in spectra with Dirac Gauginos \cite{Unwin:2012fj}, or (string-motivated) symmetries in the Higgs sector \cite{Hebecker:2012qp,Ibanez:2012zg}. Evolution of the observed Higgs quartic under renormalisation then implies $M_{\rm SUSY}(\lambda=0)\sim10^{11\pm2}$ GeV, as inferred from Standard Model-like running. This PeV-EeV mass scale is intriguing from the prospective of explaining the ``missing pulsar problem''  \cite{Bramante:2014zca,Bramante:2015dfa} and  the ``SN1a ignition problem'' \cite{Bramante:2015cua}. Moreover, there are several anomalous events observed at IceCube \cite{Aartsen:2014gkd} which have been interpreted as potential signals of the decay of superheavy dark matter \cite{Feldstein:2013kka,Esmaili:2013gha}.  

Sources of late time entropy injection commonly arise in UV complete theories, and we have emphasized that they may play a crucial role in diluting the baryon asymmetry to the observed level. Entropy dumps also provide solutions to cosmological problems related to the overproduction of stable exotics, most prominently: gravitinos \cite{Weinberg:1982zq}, axions  \cite{Preskill:1982cy,Abbott:1982af,Dine:1982ah,Steinhardt:1983ia}, axinos \cite{Rajagopal:1990yx} and GUT-monopoles \cite{Preskill:1979zi}.
We have shown that these entropy injection events can significantly change our expectation for the mass scales and couplings required for dark matter to match the observed relic density. The prospect of symmetric or asymmetric superheavy dark matter is particularly interesting given the tightening constraints on the traditional WIMP parameter space. In contrast to non-thermal models of superheavy dark matter  \cite{Chung:1998zb,Hui:1998dc,Chung:2001cb}, in this class of models the dark matter has modest couplings to Standard Model states and can be constrained by direct searches. Additionally, we have argued that for theories of high scale baryogenesis any stable state which is in thermal equilibrium with the Standard Model, and freezes-out of a radiation-dominated bath must be lighter than $10^{10}~{\rm GeV}$, allowing for maximal entropy injection after freeze-out. This limit follows from the perturbative unitarity limit \cite{Griest:1989wd} $\sigma_0 \lesssim 4 \pi /m_{\rm DM}^2$ and the maximal asymmetry bound \cite{Linde:1985gh} $\eta^{\rm initial}_{\rm B}\lesssim1$.  
The framework presented here offers new opportunities for model building, some of which are discussed above; we leave additional implementations to future publications.

\acknowledgments
We thank Chris Brust, Jakub Scholtz, and Yu-Dai Tsai for useful discussions. Research at Perimeter Institute is supported by the Government of Canada through Industry Canada and by the Province of Ontario through the Ministry of Economic Development \& Innovation.


\end{document}